\documentclass[
    twocolumn,
    english,
    reprint,
    aip,
    jcp,
    ]{revtex4-1}

\usepackage[version=3]{mhchem} 
\usepackage{pstricks}

\usepackage{amsmath,amssymb}
\usepackage{graphicx}
\usepackage{bm}
\usepackage{algorithm}         
\usepackage{algpseudocode}     
\usepackage{siunitx}

\newcommand*\latin[1]{\textit{#1}}
\newcommand*\ex[1]{\langle{}#1{}\rangle}
\newcommand*\ket[1]{|#1{}\rangle}
\newcommand*\bra[1]{\langle{}#1{}|}
\newcommand*\ad{a^{\dagger}}
\newcommand*\cd{c^{\dagger}}
\newcommand*\tr[1]{\textrm{#1}}
\newcommand*\mat[1]{\bm{#1}}
\newcommand*\mc[1]{\mathcal{#1}}
\newcommand*\ttt{\texttt}
\newcommand*\Ao{\mathring{\textrm{A}}}

\begin{document}

    \title{Incremental Embedding: A Density Matrix Embedding Scheme for Molecules}
    \author{Hong-Zhou Ye}
    \affiliation{Department of Chemistry, Massachusetts Institute of Technology, Cambridge, MA 02139}
    \author{Matthew Welborn}
    \thanks{Current address: Department of Chemistry, California Institute of Technology, Pasadena, CA 91125}
    \affiliation{Department of Chemistry, Massachusetts Institute of Technology, Cambridge, MA 02139}
    \author{Nathan D.\ Ricke}
    \affiliation{Department of Chemistry, Massachusetts Institute of Technology, Cambridge, MA 02139}
    \author{Troy \surname{Van Voorhis}}
    \email{tvan@mit.edu}
    \affiliation{Department of Chemistry, Massachusetts Institute of Technology, Cambridge, MA 02139}
    \date{\today}

    \begin{abstract}
        The idea of using fragment embedding to circumvent the high computational scaling of accurate electronic structure methods while retaining high accuracy has been a long-standing goal for quantum chemists. Traditional fragment embedding methods mainly focus on systems composed of weakly correlated parts and are insufficient when division across chemical bonds is unavoidable. Recently, density matrix embedding theory (DMET) and other methods based on the Schmidt decomposition have emerged as a fresh approach to this problem.
        Despite their success on model systems, these methods can prove difficult for realistic systems because they rely on either a rigid, non-overlapping partition of the system or a specification of some special sites (\latin{i.e.}\ ``edge" and ``center" sites), neither of which is well-defined in general for real molecules. In this work, we present a new Schmidt decomposition-based embedding scheme called \emph{Incremental Embedding} that allows the combination of arbitrary overlapping fragments without the knowledge of edge sites.
        This method forms a convergent hierarchy in the sense that higher accuracy can be obtained by using fragments involving more sites. The computational scaling for the first few levels is lower than that of most correlated wave function methods. We present results for several small molecules in atom-centered Gaussian basis sets and demonstrate that Incremental Embedding converges quickly with fragment size and recovers most static correlation in small basis sets even when truncated at the second lowest level.
    \end{abstract}

    \maketitle

    \section{Introduction}  \label{sec:introduction}

    The fast and accurate calculation of quantum mechanical properties of molecules and materials is one of the major unsolved problems in quantum chemistry. The computational cost of most accurate electronic structure methods rises sharply with system size, limiting their applications to small systems and/or moderate-sized basis sets.\cite{Bartlett07RMP,Hachmann07JCP,Chattopadhyay15JCC,Taffet17CP,Chien18JPCA,Stemmle18PRA} This scaling challenge can be potentially circumvented by fragment embedding, where the system is divided into smaller fragments, and the computationally involved, high-level theory is only required for each individual fragment. The complicated interaction between the fragment and its large-sized surroundings is then approximated by the interaction with an effective bath that mimics the rest of the system. The idea of fragment embedding serves as the basis for many methods, including fragment molecular orbital theory\cite{Kitaura99CPL,Fedorov04JCP,Khaliullin06JCP,Fedorov07JPCA} (localized molecular orbital-based embedding), subsystem density functional
    theory\cite{Senatore86PRB,Johnson87PRB,Cortona91PRB,Jacob14CMS} (density-based embedding), and dynamic mean-field theory\cite{Metzner89PRL,Georges92PRL,Georges96RMP,Maier05RMP,Turkowski12JCP} (local Green's function-based embedding) to name a few.

    A major challenge to the development of a general fragment-based method is the treatment of chemical bonds between fragments. Recently, Schmidt decomposition\cite{Klich06JPAMG,Peschel09JPAMT,Peschel12BJP} has been used for embedding fragments that are strongly correlated to a bath, which occurs when embedding fragments across chemical bonds. For each fragment, the Schmidt decomposition transform the rest of the system into an entangled, effective bath which is of \emph{same} dimension as the fragment. A low-dimensional embedding Hamiltonian is then constructed in the Schmidt space and solved accurately therein.
    In practice, a high-level calculation such as FCI (full configuration interaction\cite{Szabo96Book}), DMRG (density matrix renormalization group\cite{White92PRL,Verstraete08AIP}) or CCSD (coupled-cluster singles and doubles\cite{PurvisIII82JCP,Ahlrichs07Inbook}) is embedded in a low-level bath (usually mean-field, \latin{e.g.}\ Hartree-Fock\cite{Szabo96Book}), to recover the electron correlation missing at the mean-field level.

    In order to optimize the embedding, some matching conditions are usually imposed. So far there have been two main classes. In the first class, DMET (density matrix embedding theory\cite{Knizia12PRL,Knizia13JCTC}) and DET (density embedding theory\cite{Bulik14PRB,Bulik14JCP}), one uses rigid, non-overlapping fragments and matches the one-particle density matrix (1PDM) between the fragment and the bath. Mathematically this is achieved by adding to the low-level bath an effective one-particle potential, which changes both the low- and high-level 1PDMs. This effective potential is then tuned to satisfy the matching condition.
    This approach has shown good performance on model systems such as the Hubbard model and atomic rings/chains, even in the strong correlation domain\cite{Knizia12PRL,Knizia13JCTC,Wouters16JCTC,Zheng16PRB,Bulik14PRB,Bulik14JCP}.
    As with many fragment embedding methods, however, the restriction to non-overlapping fragments results in persistent edge effects and slow convergence with fragment size\cite{Wouters16JCTC,Zheng17PRB}. In BE (Bootstrap Embedding\cite{Welborn16JCP}), one instead uses overlapping fragments and requires in the overlapping region the ``edge" sites from one fragment agree on density matrix elements with the ``center" sites from another fragment. As long as one can make a clear distinction between the edge sites (usually on the boundary of a fragment) and the center sites (usually the most embedded part), this scheme helps to get rid of the edge effects and leads to faster convergence as demonstrated on the Hubbard model\cite{Welborn16JCP}.

    Though successful for model systems, these methods encounter difficulties with real molecules. On the one hand, the need for a rigid, non-overlapping partition of the system makes DMET/DET ambiguous when high symmetries such as translational invariance are lost. As a consequence, applications to realistic systems have been so far restricted to atomic rings\cite{Knizia13JCTC,Wouters16JCTC}, chains\cite{Knizia13JCTC} or simple polymers\cite{Bulik14JCP} in small basis sets where one atom or monomer with several basis functions can be considered as a fragment.
    One attempt at modeling real molecules was made by Wouters \latin{et al.}~\cite{Wouters16JCTC} who performed DMET calculations for the potential energy surface of a symmetric $\tr{S}_{\tr{N}}2$ reaction, \ce{C12H25F\bond{...}F^- -> C12H25F\bond{...}F^-}. However, the DMET result was less accurate than that of the full-system CCSD even with a fragment as large as four \ce{CH2} groups. BE, on the other hand, can do overlapping fragments but requires a clear definition of edge and center sites, which is also ambiguous in systems lacking certain symmetries.
    This was demonstrated by Ricke \latin{et al.}~\cite{Ricke17MP} who performed a number of BE calculations on 2D Hubbard model with fragments of different shapes and choices of center sites. According to their report, the combination that gives the best energetics is not always intuitive.

    We present here a scheme, Incremental Embedding, that allows the combination of arbitrary overlapping fragments \emph{without} the knowledge of edge sites. As a proof of concept, we test this new fragment embedding scheme on several molecules in atom-centered Gaussian basis sets. Numerical results suggest that this method converges quickly with fragment size at equilibrium geometry and recovers most static correlation in minimal basis sets, but the performance deteriorates in either bond dissociation limit or larger basis sets. We show that this arises due to the nature of the HF Schmidt bath and point out some possible solutions.

    This article is organized as follows. In Sec.\ \ref{sec:background}, we give the theoretical background through briefly reviewing the Schmidt decomposition and existing Schmidt-space fragment embedding methods. In Sec.\ \ref{sec:theory}, we formulate our theory of Incremental Embedding based on a new concept, Schmidt reduction, introduced therein. In Sec.\ \ref{sec:computational_details}, we give the computational details. Then in Sec.\ \ref{sec:results}, we present numerical results on several molecular systems as a proof of concept. In Sec.\ \ref{sec:discussion}, we discuss a potential problem of using HF as the bath. Finally in Sec.\ \ref{sec:conclusion}, we conclude this work by pointing out several future directions.

    \section{Background}    \label{sec:background}

    In the following, terminologies from lattice model are used for the formal discussion. All results can be adapted to realistic systems by replacing the ``site basis" with the appropriate one-particle basis in the corresponding scenario. For example, for molecules this could be either symmetrically orthogonalized atomic orbitals\cite{Szabo96Book} (SOAO) or localized molecular orbitals (LMO) given by some flavor of orbital localization methods (\latin{e.g.}\ the Foster-Boys scheme\cite{Boys60RMP}).

    \subsection{Schmidt Decomposition}  \label{subsec:Schmidt_decomposition}

    Suppose the system consists of two parts, the fragment (which we assume to be the minority) and the environment, such that the Hilbert space for the whole system is a direct product of the two subsystems, \latin{i.e.}\ $\mc{H} = \mc{H}_{\tr{f}} \otimes \mc{H}_{\tr{e}}$. Any state $\ket{\Psi} \in \mc{H}$ therefore has the following tensor product decomposition
    \begin{equation}    \label{eq:tensor_product_decomposition}
        \ket{\Psi}
            = \sum_{i}^{\dim{}\mc{H}_{\tr{f}}}
            \sum_{j}^{\dim{}\mc{H}_{\tr{e}}}
            \Psi_{ij} \ket{f_i} \otimes \ket{e_j},
    \end{equation}
    where $\ket{f_i} \in \mc{H}_{\tr{f}}$ and $\ket{e_j} \in \mc{H}_{\tr{e}}$ are the (many-body) basis states that span the fragment and the environment, respectively. Eqn (\ref{eq:tensor_product_decomposition}) can be brought to the Schmidt decomposed form
    \begin{equation}    \label{eq:Schmidt_decomposition_general}
        \ket{\Psi}
            = \sum_{p}^{\dim{}\mc{H}_{\tr{f}}} \lambda_p \ket{f_p} \otimes \ket{b_p}
    \end{equation}
    by a singular value decomposition on the coefficient matrix $\mat{\Psi}$, where $\ket{b_p}$'s are the so-called Schmidt entangled bath states and $\lambda_p$'s are the singular values. The benefit of the Schmidt decomposition is that the length of the expansion is limited by the number of linearly independent fragment states (those with nonzero $\lambda_p$). Thus the decomposed form has a manageable length no matter how large the original system is, as long as the fragment is not too large. Eqn (\ref{eq:Schmidt_decomposition_general}) is exact in the sense that if $\ket{\Psi}$ is a ground state of $\hat{H}$ at some level of theory, the embedding Hamiltonian
    \begin{equation}    \label{eq:Hemb_projection}
        \hat{H}_{\tr{emb}}
            = \hat{P} \hat{H} \hat{P}
    \end{equation}
    obtained by projecting $\hat{H}$ onto the Schmidt space with operator
    \begin{equation}
        \hat{P}
            = \sum_{pq}^{\dim{}\mc{H}_{\tr{f}}}
            \ket{f_p} \bra{f_p} \otimes \ket{b_q} \bra{b_q}
    \end{equation}
    shares the same ground state as $\hat{H}$ at the same level of theory. For example, the exactness of HF-in-HF embedding has been verified explicitly in Ref.~\citenum{Knizia13JCTC}.

    In electronic structure theory, the Schmidt decomposition is usually performed in the site basis. In general, for a fragment composed of $N_{\tr{frag}}$ sites $\hat{H}_{\tr{emb}}$ can be written as
    \begin{equation}    \label{eq:H_emb_general}
    \begin{split}
        \hat{H}_{\tr{emb}}
            &= \sum_{pq}^{2N_{\tr{frag}}} \tilde{h}_{pq} \cd_{p} c_{q} +
            \sum_{pqrs}^{2N_{\tr{frag}}} \tilde{V}_{pqrs} \cd_p c_q \cd_{r} c_s +   \\
            &\quad\,\sum_{pqrstu}^{2N_{\tr{frag}}} \tilde{T}_{pqrstu}
            \cd_p c_q \cd_{r} c_s \cd_t c_u + \cdots
    \end{split}
    \end{equation}
    where the summation goes over $N_{\tr{frag}}$ fragment sites and the same number of Schmidt bath sites. If one starts with a mean-field wave function, the complicated many-body Hamiltonian in eqn (\ref{eq:H_emb_general}) is truncated at the two-body level and therefore can be solved by the aforementioned accurate quantum chemical methods such as FCI, DMRG and CCSD. For this reason, nearly all Schmidt-space fragment embeddings -- including this work -- use a HF bath, with only a few exceptions\cite{Tsuchimochi15JCP}. Then in order to optimize the embedding, one often needs to impose some flavor of matching conditions. So far there have been two main classes: DMET/DET match the fragment and bath using rigid, non-overlapping fragments, and BE matches fragment to fragment when they overlap. We will review both classes in the following two subsections.

    Before that, let us put special emphasis on two potential problems underlying the HF bath. First, if some site is unentangled with all other sites in the bath wave function, it gives vanishing singular value in eqn (\ref{eq:Schmidt_decomposition_general}) and is therefore redundant for the embedding. This basis set degradation problem is in fact not rare for realistic systems with a mean-field bath, to which we will get back in Sec.\ \ref{sec:discussion}. Second, for a $K$-site, $2N$-electron system the maximum number of sites that can be entangled with each other is upper bounded by the number of electrons or holes in the system, i.e.\
    \begin{equation}    \label{eq:Maximum_number_of_entangled_sites}
        N_{\tr{frag}} \leq N_{\tr{frag}}^{\tr{max}} = \min\{N, K - N\}.
    \end{equation}
    In other words, this limits the size of fragments one can use in a Schmidt decomposition. If the system in concern is half-filled (i.e.\ $N = K$), one can expect to approach the exact solution by resorting to fragments of larger size. If this is not the case, however, such convergence is vague. In practice, a large-sized basis set is often needed in order to include as much dynamic correlation as possible. This will make the system far from half-filling and therefore deteriorate the performance of embedding with a HF bath. We will also get back to this point later in Sec.\ \ref{subsec:correlation_energies}.

    \subsection{DMET and DET}   \label{subsec:DMET_and_DET}

    In 2012, Knizia and Chan\cite{Knizia12PRL} proposed the idea to embed a high-level theory in a HF bath. To optimize the embedding, a one-particle effective potential $\hat{v}_{\tr{eff}}$ is added to the HF bath:
    \begin{equation}
        \ket{\Phi(\hat{v}_{\tr{eff}})}
            = \arg\min_{\Phi} \ex{\Phi|\hat{H} + \hat{v}_{\tr{eff}}|\Phi}
    \end{equation}
    where $\ket{\Phi}$ is restricted to be a single Slater determinant. The bath 1PDM is then made to match the fragment 1PDM by tuning this effective potential
    \begin{equation}    \label{eq:DMET_matching_condition}
        \ex{\Phi(\hat{v}_{\tr{eff}})|\cd_p c_q|\Phi(\hat{v}_{\tr{eff}})}
            = \ex{\Psi(\hat{v}_{\tr{eff}})|\cd_p c_q|\Psi(\hat{v}_{\tr{eff}})}
    \end{equation}
    where $\ket{\Psi}$ denotes the (correlated) embedding wave function, and indices $p$ and $q$ go over fragment sites only. Note that in generating $\hat{H}_{\tr{emb}}$ [eqn (\ref{eq:Hemb_projection})], $\hat{v}_{\tr{eff}}$ is involved only in the bath part. Therefore, $\ket{\Psi}$ gains its dependency on $\hat{v}_{\tr{eff}}$
    only through the bath. As demonstrated by Tsuchimochi \latin{et al.}, the matching condition in eqn (\ref{eq:DMET_matching_condition}) is not always exactly satisfiable\cite{Tsuchimochi15JCP} and therefore is optimized in a least-squares sense\cite{Wouters16JCTC}. Nevertheless, once the matching is achieved, the total energy can be expressed as a sum of fragment energies
    \begin{equation}    \label{eq:DMET_energy_general}
        E_{\tr{DMET}}
            = \sum_{A} (E_{A})_{\tr{DMET}}
            = \sum_{A} \ex{\Psi_{A}|\hat{E}_{A}|\Psi_{A}}
    \end{equation}
    where the summation goes over all (non-overlapping) fragments $\{A\}$; $\{\hat{E}_A\}$ partitions $\hat{H}$
    \begin{equation}    \label{eq:Energy_partition_over_fragments}
        \hat{H}
            = \sum_{A} \hat{E}_A
    \end{equation}
    such that each of them only involves terms that belong to the fragment as well as half the interactions between the fragment and the bath to avoid double counting. DMET is extremely powerful for strongly correlated model systems such as the Hubbard model \cite{Knizia12PRL,Zheng16PRB} and atomic rings\cite{Knizia13JCTC,Wouters16JCTC} where unique partitions of the system are obvious due to the high symmetry.
    Two years later, Bulik \latin{et al}.~ \cite{Bulik14PRB} simplified the matching condition in eqn (\ref{eq:DMET_matching_condition}) by requiring only the diagonal of the 1PDM to be matched, \latin{i.e.}\
    \begin{equation}    \label{eq:DET_matching_condition}
        \ex{\Phi(\hat{v}_{\tr{eff}})|\cd_p c_p|\Phi(\hat{v}_{\tr{eff}})}
            = \ex{\Psi(\hat{v}_{\tr{eff}})|\cd_p c_p|\Psi(\hat{v}_{\tr{eff}})}
    \end{equation}
    where $p$ goes over fragment sites only. This variant is named DET and has shown performance that are comparable with the original DMET\cite{Bulik14PRB,Bulik14JCP}.

    When applying DMET/DET to realistic systems, the absence of high symmetry makes an unambiguous non-overlapping partition of the system difficult or even impossible. Different partitions often lead to different results and there is no apparent way to evaluate the quality of those different choices. In the most general scenario, perhaps the best one can do is to adopt the following one-site embedding scheme:
    \begin{equation}    \label{eq:DMET_energy_1site}
        E_{1\tr{-site}}
            = \sum_{p}^K (E_p)_{\tr{DMET}}
    \end{equation}
    where index $p$ goes over all sites and $(E_p)_{\tr{DMET}}$ is simply the ``site energy". Eqn (\ref{eq:DMET_energy_1site}) is free of the ambiguity problem by construction, but the generalization to fragments of larger size is not obvious in the context of DMET. We will see in Sec.\ \ref{sec:theory} how this scheme could be improved systematically by Incremental Embedding.

    \subsection{Bootstrap Embedding}    \label{subsec:BootstrapEmbedding}

    In addition to the ambiguity of fragment partition, the restriction to non-overlapping fragments also results in slow convergence with fragment size due to the persistent edge effects. Recently, Welborn \latin{et al.}~\cite{Welborn16JCP} proposed the BE scheme in order to eliminate the edge effects in certain situations.
    The motivation is that when several fragments overlap, the edge sites in one fragment might be the center of another. Therefore by matching properties such as 1PDM and/or 2PDM of the former to those of the latter, one can expect improving the description of the edge sites without deteriorating the center sites. Mathematically this is formulated as a constrained optimization. Suppose fragment $A$ overlaps partially with fragment $B$, and $A$'s center sites $\mc{C}(A)$ are edge sites of $B$. Then the matching condition between $A$ and $B$ is satisfied by making the following Lagrangian stationary
    \begin{equation}    \label{eq:bootstrap_embedding_lagrangian}
    \begin{split}
        \mc{L}[\Psi_B; &\mat{\lambda}, \mat{\Lambda}]
            = \ex{\Psi_B|\hat{H}_{\tr{emb},B}|\Psi_B} + \\
            & \sum_{pq \in \mc{C}(A)} \lambda_{pq}^{B} (
            \ex{\Psi_{B}|\cd_{p}c_{q}|\Psi_{B}} - P_{pq}^{A}) + \\
            & \sum_{pqrs \in \mc{C}(A)} \Lambda_{pqrs}^{B} (
            \ex{\Psi_{B}|\cd_{p}c_{q}\cd_{r}c_{s}|\Psi_{B}} - \Gamma_{pqrs}^{A}),
    \end{split}
    \end{equation}
    which can be transformed to an eigenvalue problem of a dressed Hamiltonian
    \begin{equation}    \label{eq:Bootstrap_Embedding_Equation}
    \begin{split}
        \hat{H}^{\tr{eff}}_{\tr{emb},B}
            &= \hat{H}_{\tr{emb},B} +
            \sum_{i, j \in \mc{C}(A)} \lambda_{pq}^B \cd_p c_q + \\
            &\quad{}\sum_{pqrs \in \mc{C}(A)} \Lambda_{pqrs}^B \cd_p c_q \cd_r c_s.
    \end{split}
    \end{equation}
    In other words, the Lagrange multipliers play the role of a constraint potential $\hat{v}_{\tr{c}}$ to satisfy the matching conditions. Similar to the direct optimization method in DFT\cite{Wu03JCP,Wu05PRA}, Ricke \latin{et al}.\ proved that $\mc{L}$ has a negative semi-definite Hessian, making its stationary point a maximum and rendering eqn (\ref{eq:Bootstrap_Embedding_Equation}) numerically favorable\cite{Ricke17MP}.
    Results on model systems have suggested that this method indeed leads to faster convergence with fragment size\cite{Welborn16JCP}.

    Despite its success on model systems, the generalization of BE to a general molecular system is challenging. This is because the distinction between the edge and the center sites is often vague unless certain symmetries such as translational invariance, are present. Nevertheless, the idea of matching among overlapping fragments is of significant importance. We will see that it provides -- after being combined with the technique we are going to introduce in the next section -- a path towards realistic systems.

    \section{Theory}    \label{sec:theory}

    \subsection{Schmidt Reduction}  \label{subsec:Schmidt_Reduction}

    We first introduce a tool that enables us to encode the information from a larger embedding space to a smaller one. Suppose we start with a wave function $\ket{\Psi}$, perform the Schmidt decomposition with an $m$-site fragment, and obtain the embedding Hamiltonian $\hat{H}_{m}$ in the resulting $2m$-electron, $2m$-site space (assuming all sites are entangled). Then we solve for its ground state $\ket{\Psi_m}$, perform a second Schmidt decomposition involving $n < m$ sites, and obtain a new embedding Hamiltonian $\hat{H}_{m\to{}n}$ in the resulting $2n$-electron, $2n$-site space. Overall, the process can be summarized as
    \begin{equation}    \label{eq:Schmidt_reduction_MtoN}
        \ket{\Psi}
            \xrightarrow[\hat{H}]{m\tr{-site SD}} \hat{H}_m
            \xrightarrow{\tr{FCI}} \ket{\Psi_{m}}
            \xrightarrow[\hat{H}_m]{n\tr{-site SD}} \hat{H}_{m\to{}n}.
    \end{equation}
    We call this process a \emph{Schmidt reduction} (SR) from $m$ sites to $n$ sites, or $m \to n$ SR for short. Due to the exact nature of the Schmidt decomposition, $\hat{H}_m$ and $\hat{H}_{m \to n}$ share the same ground state even though the latter could be of much smaller dimension.

    If one starts with a HF wave function $\ket{\Phi}$, $\hat{H}_m$ has a simple form involving only one- and two-body interactions due to the mean-field nature of its bath. However, $\ket{\Psi_m}$ is correlated and so is the Schmidt bath derived from it. This renders $\hat{H}_{m \to n}$ complicated and awkward to deal with in practice [cf.\ eqn (\ref{eq:H_emb_general})]. The workaround here is to go to the Schmidt-space matrix representation.
    For instance, $\hat{H}_{m \to 1}$ can be elegantly represented by a $4$-by-$4$ matrix $\mat{H}_{m \to 1}$ in the complete one-site Schmidt basis:
    \begin{equation}    \label{eq:Schmidt_reduction_mto1_basis}
        \ket{\,}\otimes{}\ket{\uparrow\downarrow},
        \ket{\uparrow}\otimes{}\ket{\downarrow},
        \ket{\downarrow}\otimes{}\ket{\uparrow},
        \ket{\uparrow\downarrow}\otimes{}\ket{\,}.
    \end{equation}

    One of the important consequences of eqn (\ref{eq:Schmidt_reduction_MtoN}) is that it suggests an obvious way to combine overlapping fragments: reduce each of them onto the $n$ sites where they all overlap. It is then natural to require that the properties of all fragments agree on those common sites. This provides a powerful set of matching conditions that are not based on the discrimination of edge and center sites. In the following, we introduce one realization of this idea, Incremental Embedding.

    \subsection{Incremental Embedding from Two Sites to One Site} \label{subsec:Incremental_Embedding_2to1}

    One way to exploit the strength of Schmidt reduction is what we introduce in this work, \emph{Incremental Embedding} (henceforth abbreviated as IE). The goal is to construct a one-site effective Hamiltonian $\mat{H}_p$ for any given site $p$ such that it contains correlation at the level of $m$-site embeddings ($m \geq 2$). In this section, we focus on the lowest order, $m = 2$. The resulting theory is named IE from two sites to one site, or $2 \to 1$ IE for short. The generalization to $m > 2$ will be presented in Sec.\ \ref{subsec:IncEmb_mto1}.

    Suppose we have already solved the HF wave function $\ket{\Phi}$ for a system described by the following $K$-site Hamiltonian
    \begin{equation}
        \hat{H}
            = \sum_{pq}^K h_{pq} \ad_p a_q +
            \sum_{pqrs}^K V_{pqrs} \ad_p a_q \ad_r a_s.
    \end{equation}
    Then a Schmidt decomposition of $\ket{\Phi}$ on site $p$ gives the mean-field approximation to $\mat{H}_p$
    \begin{equation}
        \ket{\Phi}
            \xrightarrow[\hat{H}]{\tr{SD on }p} \mat{H}_p^0,
    \end{equation}
    which is merely the matrix representation of $\hat{H}$ in the one-site Schmidt basis of site $p$ derived from the mean-field bath. A better approximation can be obtained by the following $2 \to 1$ SR
    \begin{equation}    \label{eq:naive_IE_2to1}
        \ket{\Phi}
            \xrightarrow[\hat{H}]{\tr{SD on }(p, q)} \hat{H}_{pq}
            \xrightarrow{\tr{FCI}} \ket{\Psi_{pq}}
            \xrightarrow[\hat{H}_{pq}]{\tr{SD on } p} \mat{H}_p^{\phantom{p}q},
    \end{equation}
    where $q \neq p$ could be any other site. According to the exact nature of SR, $\mat{H}_p^{\phantom{p}q}$ contains the correlation between $p$ and $q$ at the level of two-site embedding, and thus is an improvement over $\mat{H}_p^0$. However, approximating $\mat{H}_p$ by $\mat{H}_p^{\phantom{p}q}$ is problematic: different choices of $q$ in general give different $\mat{H}_p^{\phantom{p}q}$'s and it is hard to determine which choice is better than others.

    These observations motivate one to consider how to appropriately accumulate contributions from multiple sites. A simple sum of $\mat{H}_{p}^{\phantom{p}q}$'s over several sites $q \neq p$ leads to severe double-counting problem, since each matrix alone is a representation of the full Hamiltonian $\hat{H}$. To this end, we propose to first divide the Hamiltonian into pieces, obtain the SR matrix representation for each piece using the two-site fragment that is most relevant to it, and then assemble them to construct an approximate $\mat{H}_p$ that is free of double-counting.

\begin{figure}[!h]
        \centering
        \includegraphics[width=0.95\linewidth]{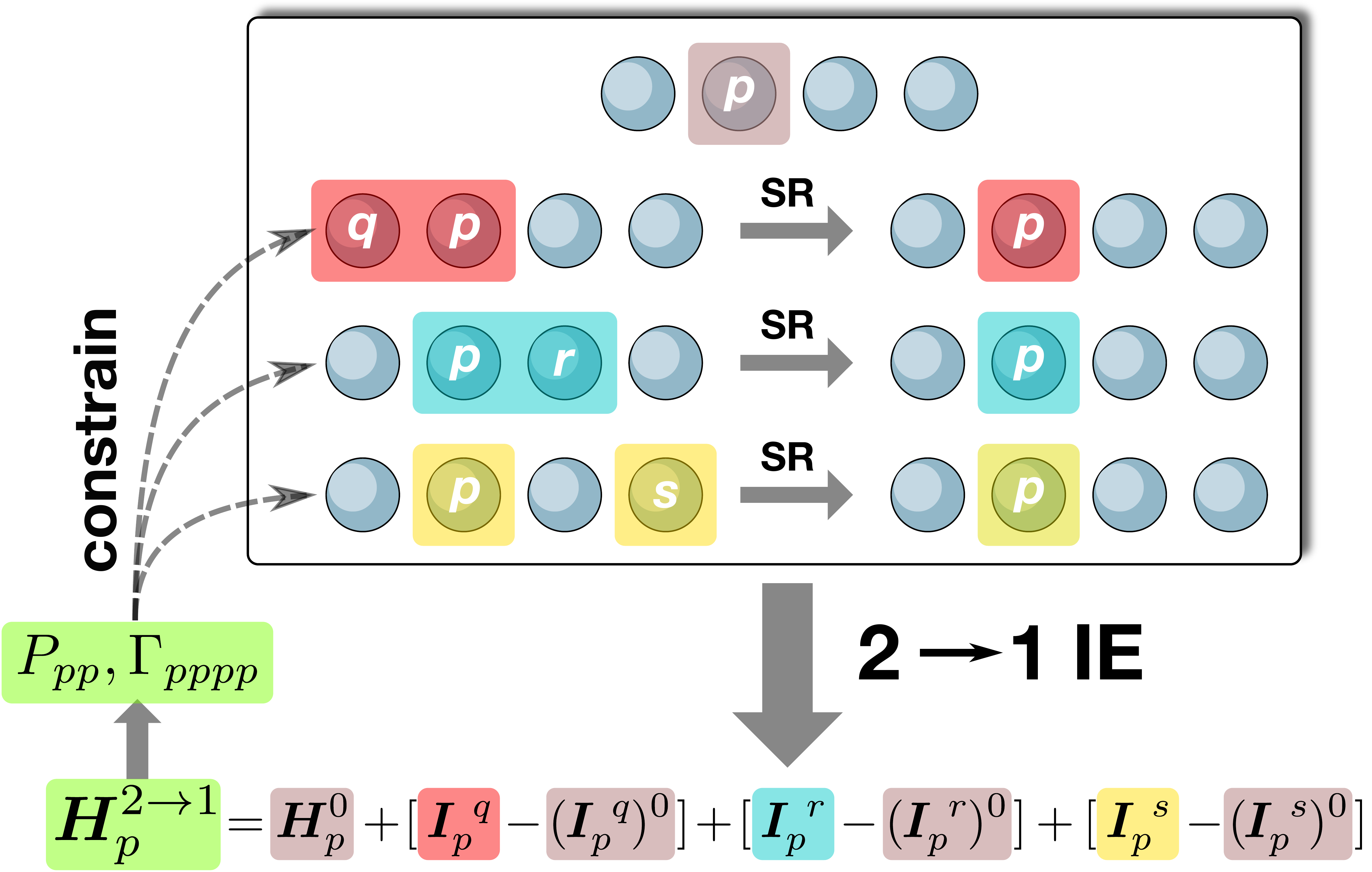}
        \caption{Schematic illustration of $2 \to 1$ IE in a four-site lattice model. The mean-field approximation $\mat{H}_p^0$ (brown) is improved by incremental Hamiltonians from three two-site embedding calculations: $(p, q)$ (red), $(p, r)$ (blue), and $(p, s)$ (yellow). The site density $P_{pp}$ and pair-density $\Gamma_{pppp}$ derived from $\mat{H}_{p}^{2 \to 1}$ (green) can in return be used to constrain the embedding calculations.}
        \label{fig:schematic_2to1}
    \end{figure}

    Specifically, consider the following partition of Hamiltonian
    \begin{equation}    \label{eq:Partition_of_H}
        \hat{H}
            = \sum_{q \neq p}^K \hat{I}_{p}^{\phantom{p}q}
    \end{equation}
    where we require each $\hat{I}_{p}^{\phantom{p}q}$ include terms that (i) belong to the fragment $(p, q)$ or (ii) are interactions between $q$ and other sites. If some term is shared by $l$ fragments, we simply attach a factor of $\frac{1}{l}$ to average it over all relevant fragments. In other words, eqn (\ref{eq:Partition_of_H}) distributes $\hat{H}$ evenly to all two-site fragments anchored by $p$. With this partition in hand, we define the \emph{incremental Hamiltonian} from two sites to one site as
    \begin{equation}    \label{eq:Incremental_Hamiltonian_2to1}
        \Delta{}\mat{I}_{p}^{2 \to 1}
            = \sum_{q \neq p}^K \mat{I}_{p}^{\phantom{p}q} -
            (\mat{I}_{p}^{\phantom{p}q})^0
    \end{equation}
    where $\mat{I}_{p}^{\phantom{p}q}$ is the matrix representation of $\hat{I}_p^{\phantom{p}q}$ in the one-site Schmidt basis of site $p$ obtained by the following $(p, q) \to p$ SR
    \begin{equation}    \label{eq:SR_for_incremental_Hamiltonian_2to1}
        \ket{\Phi}
            \xrightarrow[\hat{H}]{\tr{SD on }(p, q)} \hat{H}_{pq}
            \xrightarrow{\tr{FCI}} \ket{\Psi_{pq}}
            \xrightarrow[\hat{I}_{p}^{\phantom{p}q}]{\tr{SD on }p} \mat{I}_{p}^{\phantom{p}q}
    \end{equation}
    and $(\mat{I}_{p}^{\phantom{p}q})^0$ is its mean-field counterpart
    \begin{equation}
        \ket{\Phi}
            \xrightarrow[\hat{I}_{p}^{\phantom{p}q}]{\tr{SD on }p} (\mat{I}_{p}^{\phantom{p}q})^0.
    \end{equation}
    The physical meaning of $\Delta{}\mat{I}_{p}^{2 \to 1}$ is clear: it accumulates the correlations between site $p$ and all other sites that are missing at the mean-field level. Adding this correction to $\mat{H}_p^0$, we have a better approximation to $\mat{H}_p^0$,
    \begin{equation}    \label{eq:Incremental_Embedding_2to1}
        \mat{H}_p^{2 \to 1}
            = \mat{H}_p^0 + \Delta{}\mat{I}_p^{2 \to 1}.
    \end{equation}
    One can expect $\mat{H}_p^{2 \to 1}$ to be of the quality of two-site embeddings, since each piece of the Hamiltonian is improved by the embedding calculation involving the most relevant two-site fragment. This is also schematically illustrated in FIG.\ \ref{fig:schematic_2to1}.

    Once we obtain the effective Hamiltonians $\{\mat{H}_{p}^{2 \to 1}\}$ for all sites, we can readily determine their ground states $\{\bm{u}_p^{2 \to 1}\}$ (as the eigenvectors of the lowest eigenvalue) and compute the site densities $\{P_{pp}\}$ (\latin{vide infra}). In general, they do not add up to the correct number of electrons, because each $\mat{H}_{p}^{2 \to 1}$ is generated from multiple fragments of different chemical potentials. This violation in the conservation of particle number can be fixed by introducing a global chemical potential $\mu$, which is determined by solving
    \begin{equation}
        \sum_{p}^{K} P_{pp}(\mu)
            = N
    \end{equation}
    where the $\mu$-dependent site densities are obtained from solving $\{\mat{H}_{p}^{2 \to 1} + \mu\mat{D}\}$ (as opposed to the bare Hamiltonians); $\mat{D} = \tr{diag}\,(0, 1, 1, 2)$ in the Schmidt basis shown in eqn (\ref{eq:Schmidt_reduction_mto1_basis}).
    We note that it is also possible to apply a set of site-specific chemical potentials $\{\mu_p\}$ to tune the population for each site. This is useful when IE is performed in a non-self-consistent manner (see Sec.\ \ref{subsec:density_optimization}).

    \subsection{Expectation Values in $2 \to 1$ IE} \label{subsec:expectation_values_2to1}

    In last section, we discussed how a one-site effective Hamiltonian can be constructed from successively improving the mean-field description by incorporating the correlation with every other site. Once done, the ground state for each site is approximated by a four-dimensional vector, $\{\bm{u}_p^{2 \to 1}\}$, in the one-site Schmidt basis [eqn (\ref{eq:Schmidt_reduction_mto1_basis})]. The ground state expectation value of any given operator can then be evaluated by summing contributions from each site. Let us take the total energy as an example, whose corresponding operator is the Hamiltonian $\hat{H}$. First we obtain a partition of $\hat{H}$ over all sites
    \begin{equation}    \label{eq:Energy_partition_over_sites}
        \hat{H}
            = \sum_p^{K} \hat{E}_p
    \end{equation}
    which is a special case of eqn (\ref{eq:Energy_partition_over_fragments}) with each fragment involving only one site. For each site $p$, we further partition $\hat{E}_p$ by distributing it evenly to all relevant two-site fragments in a way similar to eqn (\ref{eq:Partition_of_H})
    \begin{equation}    \label{eq:Energy_partition_per_site}
        \hat{E}_p
            = \sum_{q \neq p}^K \hat{E}_{p}^{\phantom{p}q}.
    \end{equation}
    With this partition, a matrix representation of $\hat{E}_p$ in the Schmidt basis of $2 \to 1$ IE can be constructed by the same procedure described in eqns (\ref{eq:Incremental_Hamiltonian_2to1} -- \ref{eq:Incremental_Embedding_2to1})
    \begin{equation}    \label{eq:Incremental_Embedding_energy_matrix_2to1}
        \mat{E}_p^{2 \to 1}
            = \mat{E}_p^{0} + \Delta{}\mat{E}_p^{2 \to 1},
    \end{equation}
    which gives the site energy for $p$ at the level of $2 \to 1$ IE
    \begin{equation}    \label{eq:Incremental_Embedding_site_energy_2to1}
        E_p^{2 \to 1}
            = (\bm{u}_p^{2 \to 1})^{\dagger} \mat{E}_p^{2 \to 1}
            \bm{u}_p^{2 \to 1}.
    \end{equation}
    Finally, the total energy for $2 \to 1$ IE is simply a sum of all site energies
    \begin{equation}    \label{eq:Incremental_Embedding_energy_2to1}
        E^{2 \to 1}
            = \sum_{p}^K E_p^{2 \to 1}.
    \end{equation}
    Eqn (\ref{eq:Incremental_Embedding_energy_2to1}) can be viewed as an unambiguous generalization of the one-site DMET energy in eqn (\ref{eq:DMET_energy_1site}) to two-site fragments. The generalization to an arbitrary number of fragment sites is presented in the following section.

    Note that the process above becomes extremely simple, if the operator involves only one site, say $p$. In that scenario, one can bypass the partition and summation steps [eqn (\ref{eq:Energy_partition_over_sites} -- \ref{eq:Incremental_Embedding_energy_matrix_2to1})], and obtain the matrix representation of that operator using any fragments involving $p$.
    Examples of this type include the diagonal elements of 1PDM (site densities, $\{P_{pp}\}$) and 2PDM (pair-densities, $\{\Gamma_{pppp}\}$).

    \subsection{Generalization to Fragments of Arbitrary Size}    \label{subsec:IncEmb_mto1}

    In this section, we generalize IE to fragments of arbitrary size, in a way that is similar to the method of increments commonly used in local correlation methods\cite{Pulay83CPL,Pulay86TCA,Saebo87JCP,Hampel96JCP}.
    The end results can be summarized in the following recursive formula for $m \geq 3$
    \begin{equation}    \label{eq:Incremental_Embedding_mto1}
        \mat{H}_p^{m \to 1}
            = \mat{H}_{p}^{(m - 1) \to 1} + c_{m \to 1}
            \Delta{}\mat{I}_{p}^{m \to 1}
    \end{equation}
    where $c_{m \to 1}$ is an appropriate constant that ensures the series of equations terminate appropriately (\latin{vide infra}); $\Delta{}\mat{I}_{p}^{m \to 1}$ is the incremental Hamiltonian from $m$ sites to one site,
    \begin{equation}    \label{eq:Incremental_Hamiltonian_mto1}
    \begin{split}
        \Delta{}\mat{I}_{p}^{3 \to 1}
            &= \sum_{r>q\neq p}^K \mat{I}_{p}^{\phantom{p}qr} -
            \mat{I}_{p}^{\phantom{p}q} - \mat{I}_{p}^{\phantom{p}r} \\
        \Delta{}\mat{I}_{p}^{4 \to 1}
            &= \sum_{s>r>q\neq p}^K \mat{I}_{p}^{\phantom{p}qrs} -
            \mat{I}_{p}^{\phantom{p}qr} - \mat{I}_{p}^{\phantom{p}rs} \\
            \phantom{=}& - \mat{I}_{p}^{\phantom{p}qs} +
            \mat{I}_{p}^{\phantom{p}p} +
            \mat{I}_{p}^{\phantom{p}r} + \mat{I}_{p}^{\phantom{p}s}\\
        &\cdots
    \end{split}
    \end{equation}
    where terms like $\mat{I}_p^{\phantom{p}qr}$ and $\mat{I}_p^{\phantom{p}qrs}$ are matrix representations of the sum of relevant pieces of $\hat{H}$ as defined in eqn (\ref{eq:Partition_of_H}). For example, $\mat{I}_p^{\phantom{p}qr}$ can be obtained by the following $3 \to 1$ SR
    \begin{equation}    \label{eq:Define_Ipqr}
        \ket{\Phi}
            \xrightarrow[\hat{H}]{\tr{SD on }(p,q,r)} \hat{H}_{pqr}
            \xrightarrow{\tr{FCI}} \ket{\Psi_{pqr}}
            \xrightarrow[\hat{I}_p^{\phantom{p}q} + \hat{I}_p^{\phantom{p}r}]{\tr{SD on }p}\mat{I}_p^{\phantom{p}qr}
    \end{equation}
    The physical meaning of $\Delta{}\mat{I}_p^{m \to 1}$ is also straightforward: it is the correction from $m$-site embedding calculations that are not included in any $(m-1)$-site embedding calculations.

    The constant coefficients $\{c_{m \to 1}\}$ arise due to the difference between traditional incremental methods (such as the aforementioned local correlation methods) and IE. In local correlation methods, a hierarchy similar to eqn (\ref{eq:Incremental_Embedding_mto1}), but with $c_{m \to 1} \equiv 1$ for all $m$, can be derived, which terminates when all sites are involved (\latin{i.e.}\ $m = K$). In that situation, the local correlation method is exact in the sense that it is equivalent to applying the same correlation method to all sites. In IE, on the other hand, the highest level one can go with eqn (\ref{eq:Incremental_Embedding_mto1}) is limited by the maximum number of entangled sites, $N_{\tr{frag}}^{\tr{max}}$ [eqn (\ref{eq:Maximum_number_of_entangled_sites})].
    If one requires that the highest level of IE be
    \begin{enumerate}
        \item[(i)] exact when the system is at half-filling (\latin{i.e.}\ $N_{\tr{frag}}^{\tr{max}} = K/2$), and
        \item[(ii)] an average of all $N_{\tr{frag}}^{\tr{max}}$-site embedding calculations otherwise,
    \end{enumerate}
    the following expressions for $\{c_{m \to 1}\}$ can be derived
    \begin{equation}    \label{eq:Incremental_Embedding_normalization_coeff}
        c_{m \to 1}
            = {K - m \choose N_{\tr{frag}}^{\tr{max}} - m} \bigg/
            {K - 2 \choose N_{\tr{frag}}^{\tr{max}} - 2}.
    \end{equation}
    Note that when $N_{\tr{frag}}^{\tr{max}} = K$, eqn (\ref{eq:Incremental_Embedding_normalization_coeff}) gives $c_{m \to 1} \equiv 1$ and hence formally reduces to traditional local correlation methods.
    Moreover, eqn (\ref{eq:Incremental_Embedding_normalization_coeff}) gives $c_{2 \to 1} = 1$ for $m = 2$, which is also consistent with $2 \to 1$ IE [eqn (\ref{eq:Incremental_Embedding_2to1})].

    In order to generalize the energy evaluation scheme, we need to generalize eqn (\ref{eq:Incremental_Embedding_energy_matrix_2to1}) to multiple sites. To that end, a recursive formula similar to eqn (\ref{eq:Incremental_Embedding_mto1}) can be derived for $\mat{E}_p$,
    \begin{equation}
        \mat{E}_p^{m \to 1}
            = \mat{E}_p^{(m-1) \to 1} + c_{m \to 1} \Delta{}\mat{E}_{p}^{m \to 1},
    \end{equation}
    where $\{c_{m \to 1}\}$ is the same set of coefficients given by eqn (\ref{eq:Incremental_Embedding_normalization_coeff}). With this in hand, the site energy
    \begin{equation}
        E_p^{m \to 1}
            = (\bm{u}_p^{m \to 1})^{\dagger} \mat{E}_p^{m \to 1}
            \bm{u}_p^{m \to 1},
    \end{equation}
    and the total energy
    \begin{equation}    \label{eq:Incremental_Embedding_energy_mto1}
        E^{m \to 1}
            = \sum_{p}^K E_p^{m \to 1}
    \end{equation}
    for $m \to 1$ IE can be straightforwardly evaluated, where $\bm{u}_p^{m \to 1}$ is the lowest eigenvector of $\mat{H}_p^{m \to 1}$ (with a proper chemical potential). Eqn (\ref{eq:Incremental_Embedding_energy_mto1}) can be viewed as an unambiguous generalization of the one-site DMET energy in eqn (\ref{eq:DMET_energy_1site}) to fragments composed of an arbitrary number of sites.

    \subsection{Matching Conditions}    \label{subsec:matching_conditions}

    So far we have not touched one of the most powerful ingredients in embedding calculations -- the matching condition. From the discussion above, constructing $\mat{H}_{p}^{m \to 1}$ requires embedding calculations for all $m$-site fragments involving site $p$. Without any constraints, these overlapping fragments in general will not agree with one another on their common site, with the only exception where the exact bath is used as opposed to the mean-field approximation. This observation indicates that one can optimize these embedding calculations by forcing the match to happen.

    Suppose we have obtained $\{\mat{H}_p^{m \to 1}\}$ for all sites. Solving them under an appropriate chemical potential, we can compute $\{P_{pp}\}$ and $\{\Gamma_{pppp}\}$ as described in Sec.\ \ref{subsec:expectation_values_2to1}. These values are our current best estimation of the diagonal elements of the exact 1PDM and 2PDM. Naturally, we can require the site densities and pair-densities of all fragments match them. Mathematically, the problem of constraining certain density matrix elements to given values can be formulated as a constrained optimization, and has already been addressed in BE [eqn (\ref{eq:bootstrap_embedding_lagrangian}) and (\ref{eq:Bootstrap_Embedding_Equation})].\cite{Welborn16JCP,Ricke17MP}
    Here, we adapt the method to IE. Suppose we want to constrain both site densities and pair-densities for a two-site fragment $(p, q)$. We can achieve this by introducing the following Lagrangian
    \begin{equation}    \label{eq:Lagrangian_2to1}
    \begin{split}
        \mc{L}_{pq}[\Psi_{pq}; &\bm{\lambda}, \bm{\Lambda}]
            = \ex{\hat{H}_{pq}}_{pq} +                        \\
            &\sum_{r = p, q} \big[ \lambda_{r}(\ex{\ad_r a_r}_{pq} - P_{rr}) +   \\
            &\Lambda_{r}(\ex{\ad_r a_r \ad_r a_r}_{pq} - \Gamma_{rrrr}) \big]
    \end{split}
    \end{equation}
    where $\hat{H}_{pq}$ is the embedding Hamiltonian; $\{P_{rr}\}$ and $\{\Gamma_{rrrr}\}$ are the target values; $\ex{\cdots}_{pq}$ is short for $\ex{\Psi_{pq}|\cdots|\Psi_{pq}}$.
    Making $\mc{L}_{pq}$ stationary leads to the following eigenvalue equation
    \begin{equation}    \label{eq:dressed_eigenvalue_problem_2to1}
        (\hat{H}_{pq} + \hat{v}^{\tr{c}}_{pq}) \ket{\Psi_{pq}}
            = \mc{E}_{pq} \ket{\Psi_{pq}}
    \end{equation}
    where $\hat{H}_{pq}$ is dressed by a constraint potential
    \begin{equation}    \label{eq:constraint_potential_2to1}
        \hat{v}^{\tr{c}}_{pq}
            = \sum_{r = p, q} (\lambda_r \ad_r a_r +
            \Lambda_r \ad_r a_r \ad_r a_r).
    \end{equation}
    Eqns (\ref{eq:dressed_eigenvalue_problem_2to1}) and (\ref{eq:constraint_potential_2to1}) enable us to apply the desired constraints to fragment calculations readily in a ground state formalism. Note that the constraint potential only exists in obtaining $\ket{\Psi_{pq}}$, and should not be included in other steps of SR.

    \subsection{Density Optimization}   \label{subsec:density_optimization}

    Once the matching conditions are imposed to each fragment in IE, one can construct a new set of $\{\mat{H}_p^{m \to 1}\}$ and recompute the site densities and pair-densities. In general, these values are of better quality compared to the old estimation, due to the embedding being optimized by the matching conditions. In return, these new densities can be used to constrain further embedding calculations which will generate $\{\mat{H}_p^{m \to 1}\}$ of even better quality. This process can be repeated until self-consistency is reached, making the theory a closed loop (FIG.\ \ref{fig:schematic_2to1}).

    The discussion above immediately suggests an algorithm to optimize the densities in IE self-consistently. Here, we state it for $2 \to 1$ IE for the sake of simplicity, and the generalization to larger fragments should be straightforward.
    \begin{enumerate}
        \item Solve the HF wave function $\ket{\Phi}$ for the whole system; obtain the mean-field Hamiltonians $\{\mat{H}_p^{0}\}$ for all sites.
        \item Obtain some guess densities $\{P^{(0)}_{pp}\}$ and $\{\Gamma^{(0)}_{pppp}\}$ (e.g.\ HF).  \label{step:init_guess}
        \item Perform embedding calculations for all two-site fragments; in each calculation, constrain the site densities and pair-densities of the fragment sites to match $\{P^{(0)}_{pp}\}$ and $\{\Gamma^{(0)}_{pppp}\}$, respectively.
        \item For each site $p$, Schmidt reduce all $p$-involved fragments to $p$, and compute $\mat{H}_{p}^{2 \to 1}$ according to eqn (\ref{eq:Incremental_Embedding_2to1}).
        \item Diagonalize $\{\mat{H}_p^{2 \to 1}\}$ under an appropriate global chemical potential $\mu$; obtain the ground states $\{\bm{u}_p^{2 \to 1}\}$, and recompute $\{P_{pp}\}$ and $\{\Gamma_{pppp}\}$. \label{step:chemical_potential}
        \item If the new densities do not match $\{P^{(0)}_{pp}\}$ and $\{\Gamma^{(0)}_{pppp}\}$, go back to step \ref{step:init_guess} with the new guess densities; otherwise, the density optimization is converged, and the ground state expectation values of desired operators can be computed using $\{\bm{u}_p^{2 \to 1}\}$.
    \end{enumerate}

    In addition to the self-consistent version, we note that IE can also be formulated as a non-self-consistent theory. In terms of the algorithm, the main difference lies in step \ref{step:chemical_potential}: instead of solving for a global chemical potential, one determines a set of site-specific chemical potentials $\{\mu_p\}$, such that for each site the population matches the guess densities. In other words, the densities are not optimized and IE is used in a one-shot style, similar to the $G_0W_0$ method.\cite{Hybertsen86PRB,Godby88PRB,vanSchilfgaarde06PRL}
    This approximation would be useful when (i) the system is large and hence full self-consistency is expensive, and (ii) the quality of the guess densities is reasonable. We will examine the performance of this approximation in Sec.\ \ref{sec:results}.

    \subsection{Computational Scaling}  \label{subsec:computational_scaling}

    We end this section by briefly discussing the computational scaling. For $m \to 1$ IE, the total work is dominated by the embedding calculations of ${K \choose m} \sim O(K^m)$ $m$-site fragments. Symmetries can effectively reduce this number by a constant factor, as fragments related by symmetry operations will give same one-site Hamiltonian and hence need to be evaluated only once. For each fragment, there are two potential rate-limiting steps: (i) the basis transformation of integrals (including the partitioned Hamiltonian $\{\hat{I}_p^{\phantom{p}q}\}$) from the site basis to the Schmidt basis, and (ii) the high-level calculation (in this paper, FCI).
    For small fragments ($m \ll K$), the former dominates with a $O(K^4)$ scaling, which will exceed the aforementioned $O(K^m)$ scaling if $m < 4$. Fortunately, the basis transformation needs to be performed only once, due to the mean-field bath not being optimized in IE. This feature makes the basis transformation step usually negligible, especially in the self-consistent version. Under these conditions, $m \to 1$ IE has a scaling of $O(K^m)$, which is lower than most accurate quantum chemical methods [$> O(K^5)$] if the incremental expansion [eqn (\ref{eq:Incremental_Embedding_mto1})] can be truncated at a small $m$.

    \section{Computational Details} \label{sec:computational_details}

    In the following computational work, we will examine the performance of IE using several small molecules. The symmetrically orthogonalized atomic orbitals\cite{Szabo96Book} (SOAOs) are used as site basis for the radial expansion of the hydrogen ring, while localized molecular orbitals (LMOs) of the Foster-Boys style\cite{Boys60RMP} are used in all other cases. The necessary atomic integrals are generated by \ttt{Psi4}\cite{Parrish17JCTC}. The Foster-Boys localization is performed in Q-Chem\cite{Shao15MP}. Molecular geometries are also optimized in Q-Chem at the B3LYP\cite{Becke93JCP}/cc-pVTZ\cite{Dunning89JCP} level and can be found in Supporting Information. All embedding calculations, including IE, BE and one-site DMET, are performed using the electronic structure program, \ttt{frankenstein},\cite{Ye17Github} developed by one of the authors.
    Spin-restricted HF (RHF) and FCI are used as bath and high-level solvers, respectively. In self-consistent IE, both site densities and pair-densities are self-consistently determined based on an RHF guess. In the non-self-consistent version, only the site densities are constrained to the RHF values due to the bad quality of the mean-field pair-densities. For one-site DMET, we abandon the self-consistency and also constrain the site densities to the RHF values. For all systems tested in this work, the exact solutions are accessible and obtained by the \ttt{Block} DMRG code\cite{Chan02JCP,Chan04JCP,Ghosh08JCP,Sharma12JCP,OlivaresAmaya15JCP}.

    \section{Results}   \label{sec:results}

    \subsection{Radial Expansion of the Hydrogen Ring Model}    \label{subsec:hydrogen_ring}

    We select the minimal-basis hydrogen ring model as our first example for several reasons. First it can be viewed as the simplest generalization of the Hubbard model towards real molecules, covering both weakly correlated domain (near equilibrium geometry) and strongly correlated domain (dissociation limit). Second, it is an ``easy" case for Schmidt-space embedding with a HF bath according to our discussion in Sec.\ \ref{subsec:Schmidt_decomposition} because the system is at half-filling. Last but not least, the cyclic symmetry makes all sites (which are symmetrically orthogonalized $1s$ orbitals in this case) equivalent. This not only renders BE applicable for comparison, but also tremendously reduces the computational work for IE so that the trend of convergence with fragment size can be examined thoroughly.

    \begin{figure}[!h]
        \centering
        \includegraphics[width=1.00\linewidth]{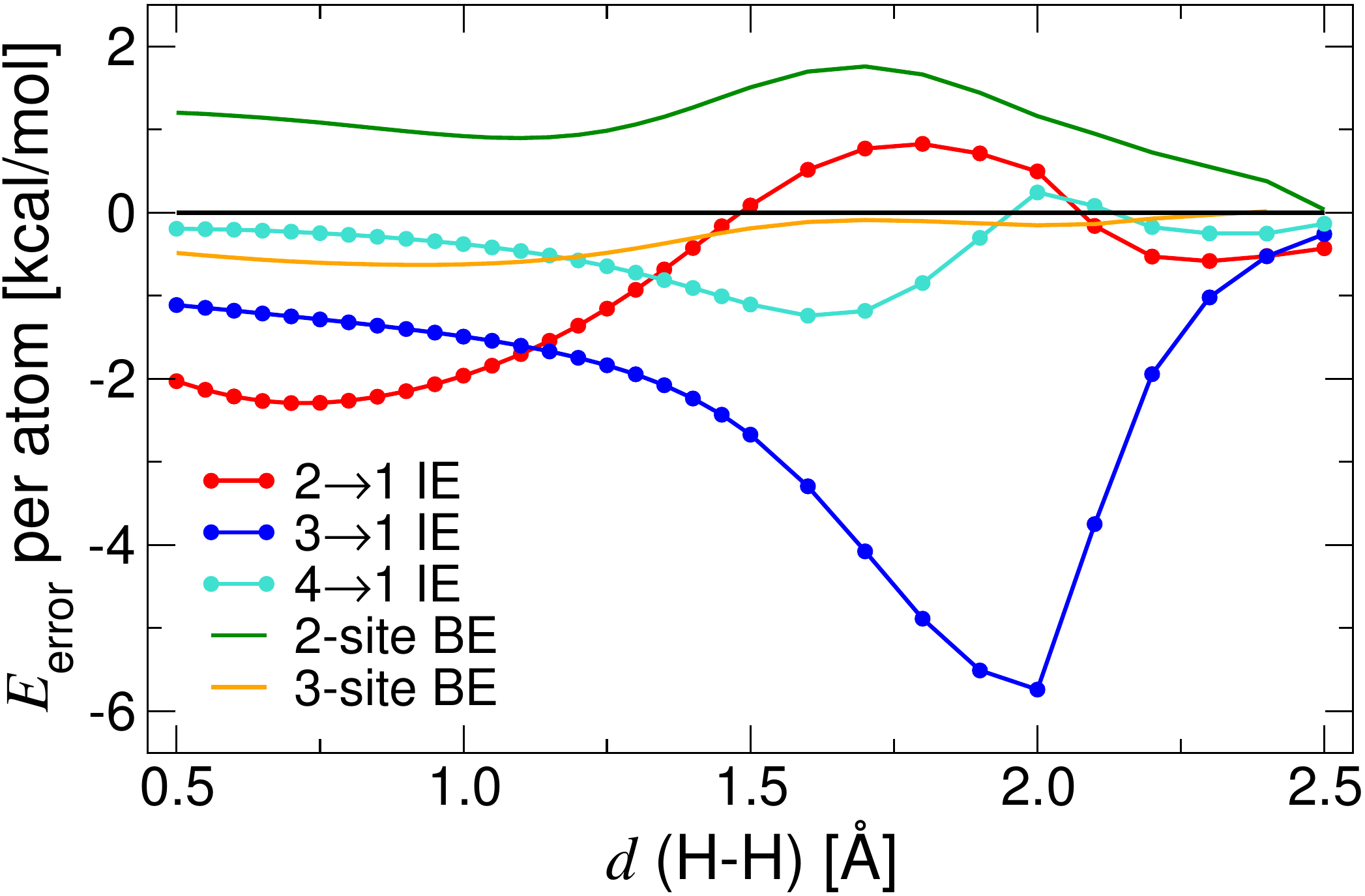}
        \caption{Total energy error per atom (in kcal/mol) of the radial expansion of STO-3G \ce{H10}.}
        \label{fig:h10}
    \end{figure}

    In FIG.\ \ref{fig:h10} the energy errors per atom in the radial expansion of STO-3G \ce{H10} are plotted for IE and BE, respectively.
    The non-self-consistent version of IE is used since the site densities are completely determined by the cyclic symmetry. For BE, fragments involving two and three adjacent sites are used, but only in the latter is there the distinction of center and edge sites. In that case, the pair-densities of edge sites are made to match that of the center site. Due to the half-filled configuration, both methods become exact when using fragments composed of five sites. Overall, both methods are very accurate even at the $2$-site level. The error is consistently small ($<2$ kcal/mol per atom) for all geometries tested here. Near the equilibrium position ($\sim 0.95$ $\Ao$), systematic improvements are observed for both methods when fragments of larger size are used. In the dissociation limit, correct asymptotic behavior is recovered in all cases even though the HF bath is spin-restricted.
    At intermediate geometries ($1.5 \sim 2.0$ $\Ao$), however, the convergence with fragment size is not monotonic for IE: the $3 \to 1$ level suffers from severe over-correlation and is worse than $2 \to 1$; this over-correlation is only ameliorated by corrections from the $4 \to 1$ level. Quite the contrary, BE continues to reduce the error by using a larger fragment and shows better accuracy compared to IE with the same fragment size. Nevertheless, the performance loss of IE compared to BE is somewhat expected in this specific example, since the model is Hubbard-like and therefore optimal for the latter.

    \subsection{Single Bond Breaking} \label{subsec:single_bond_breaking}

    \begin{figure*}[!h]
        \centering
        \includegraphics[width=0.48\linewidth]{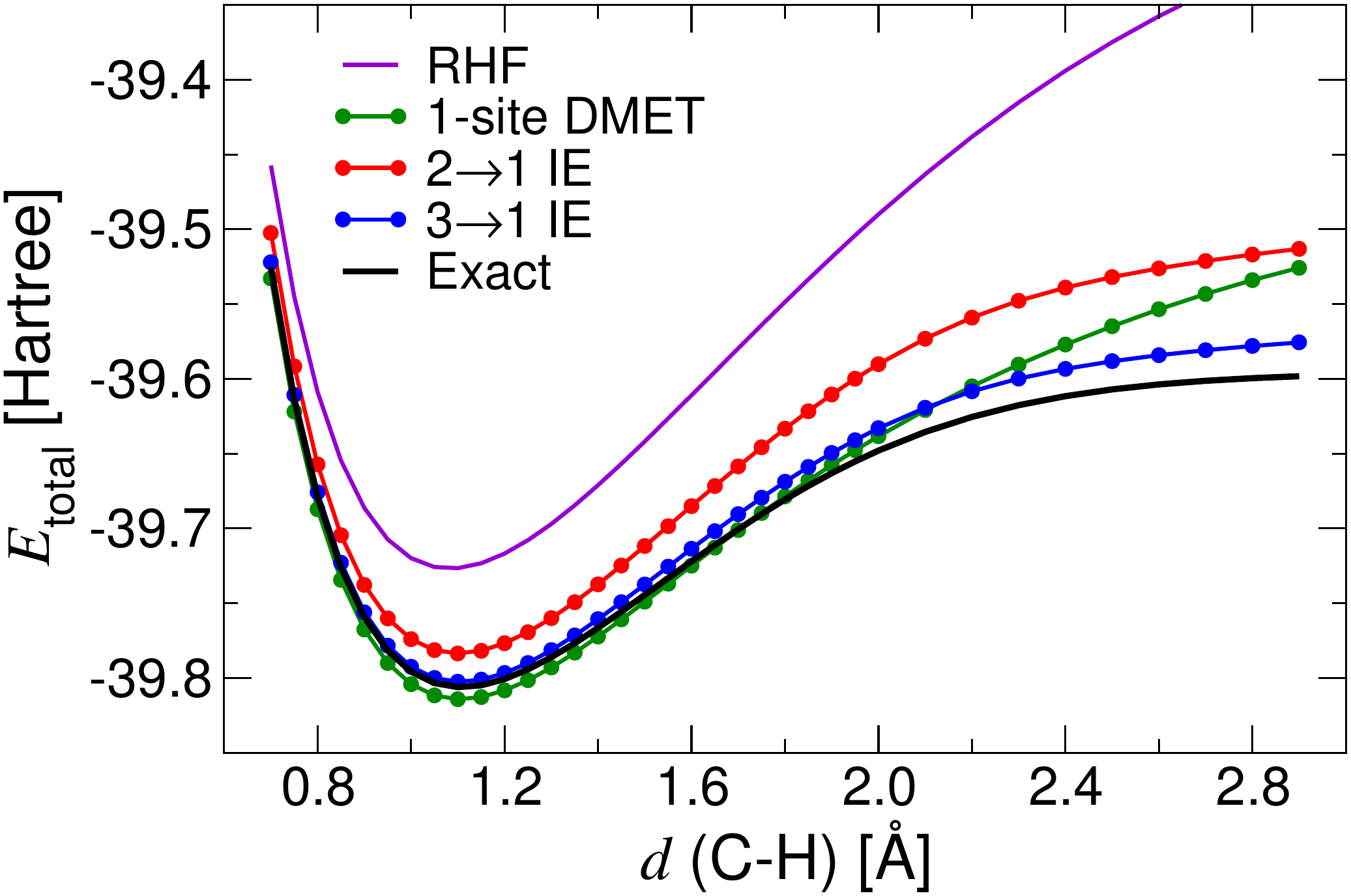}\hfill
        \includegraphics[width=0.48\linewidth]{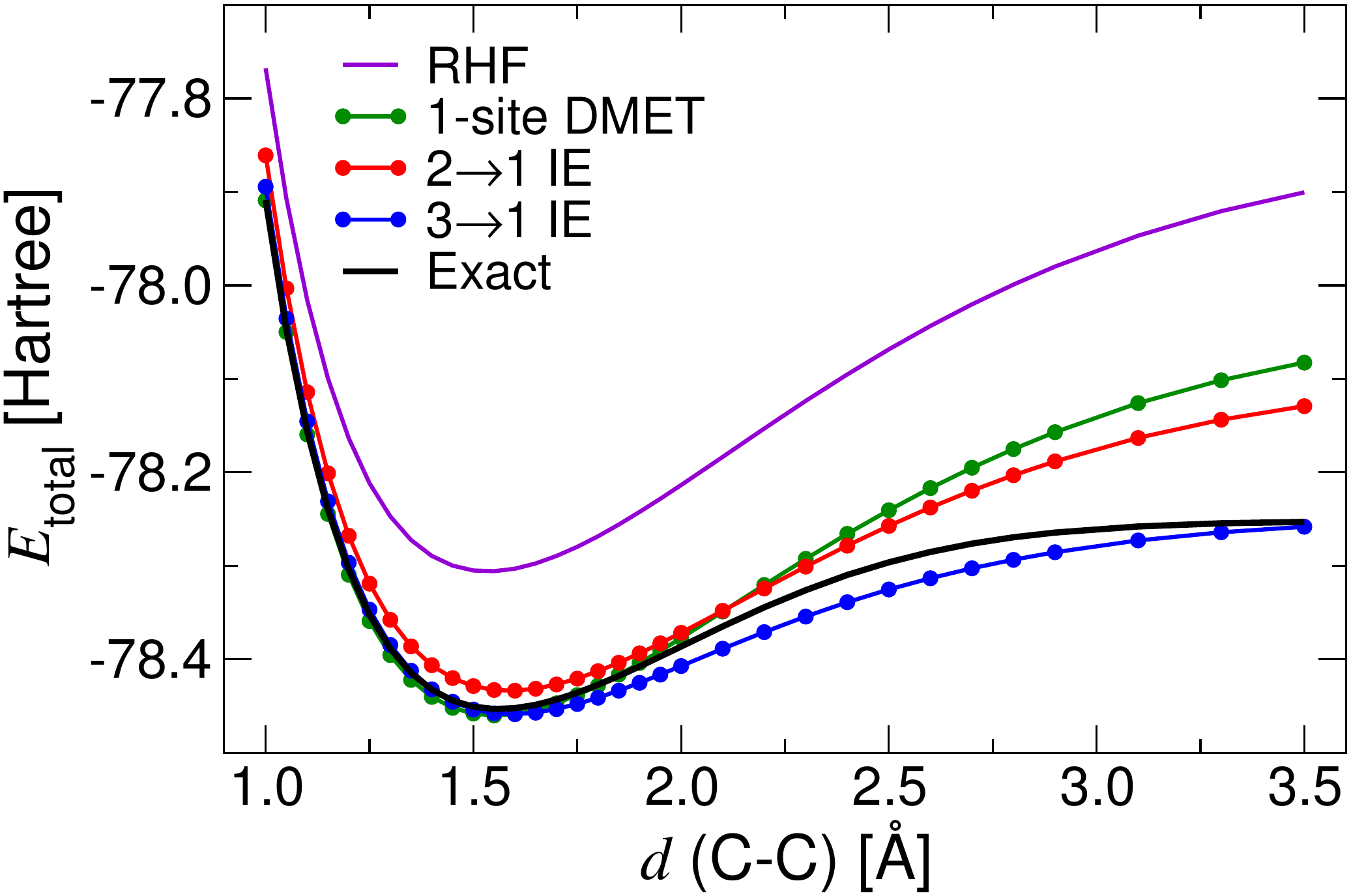}
        \caption{Potential energy surfaces of (a) stretching one \ce{C-H} bond of a methane molecule and (b) symmetrically dissociating an ethane molecule into two methyl radicals, predicted by the non-self-consistent IE and one-site DMET. The STO-3G basis set is used in both cases.}
        \label{fig:c2h6_nsc}
    \end{figure*}

    Now we consider two real molecules, \ce{CH4} and \ce{C2H6} in their minimal basis (STO-3G), to which BE is no longer applicable. Both molecules show merely a small deviation from the ideal half-filled configuration. Therefore, we can still expect good performance from the embedding calculations. Specifically we are interested in the energetics of the following two single-bond breaking processes:
    \begin{equation}
    \begin{split}
        &\ce{CH4 -> CH3. + H.}   \\
        &\ce{C2H6 -> CH3. + CH3.}.
    \end{split}
    \end{equation}
    First we consider the PESs obtained by non-self-consistent IE as shown in FIG.\ \ref{fig:c2h6_nsc}, along with non-self-consistent, one-site DMET results for comparison. In both methods, the site densities are constrained to the RHF values, whose quality is high near equilibrium geometry but deteriorates quickly as the bond is stretched (see Figure S1 in Supporting Information). If the error is mainly density-driven, we should expect good accuracy at equilibrium geometries as well as increasing error along the dissociation processes. This is indeed the case for one-site DMET (green), as can be seen from the growing gap between the embedding solution and the exact one in FIG.\ \ref{fig:c2h6_nsc}. $2 \to 1$ IE (red) shows similar trend when approaching the dissociation limit, but recovers only a limited amount of correlation energies at equilibrium position. On the contrary, $3 \to 1$ IE (blue) predicts equilibrium energies of very high accuracy (almost overlap the exact solution) and also improves the asymptotic behaviors significantly. This indicates that IE could effectively mitigate the sensitivity to the quality of the underlying approximate densities by using fragments of larger size.

    \begin{figure*}[!h]
        \centering
        \includegraphics[width=0.48\linewidth]{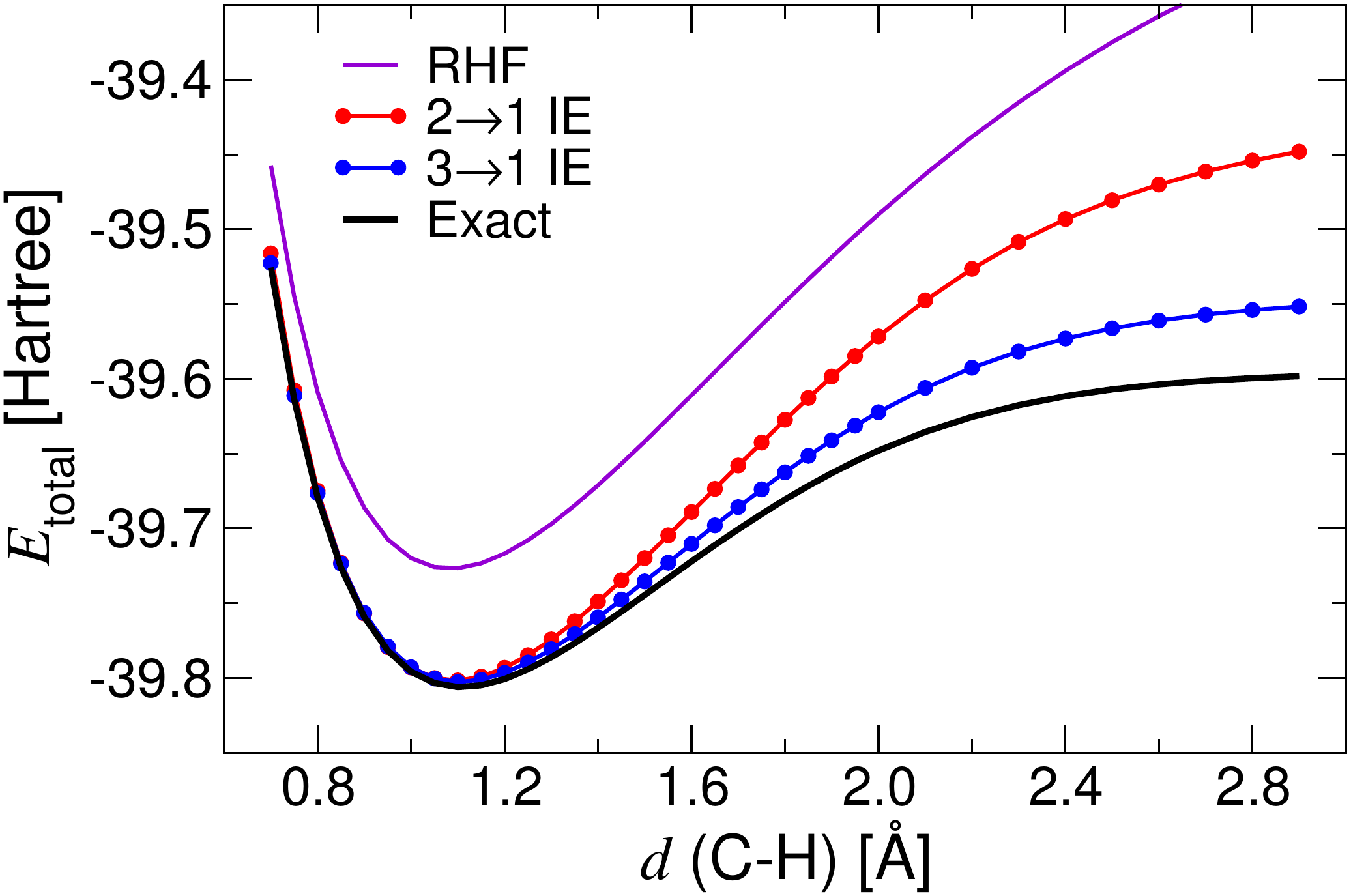}\hfill
        \includegraphics[width=0.48\linewidth]{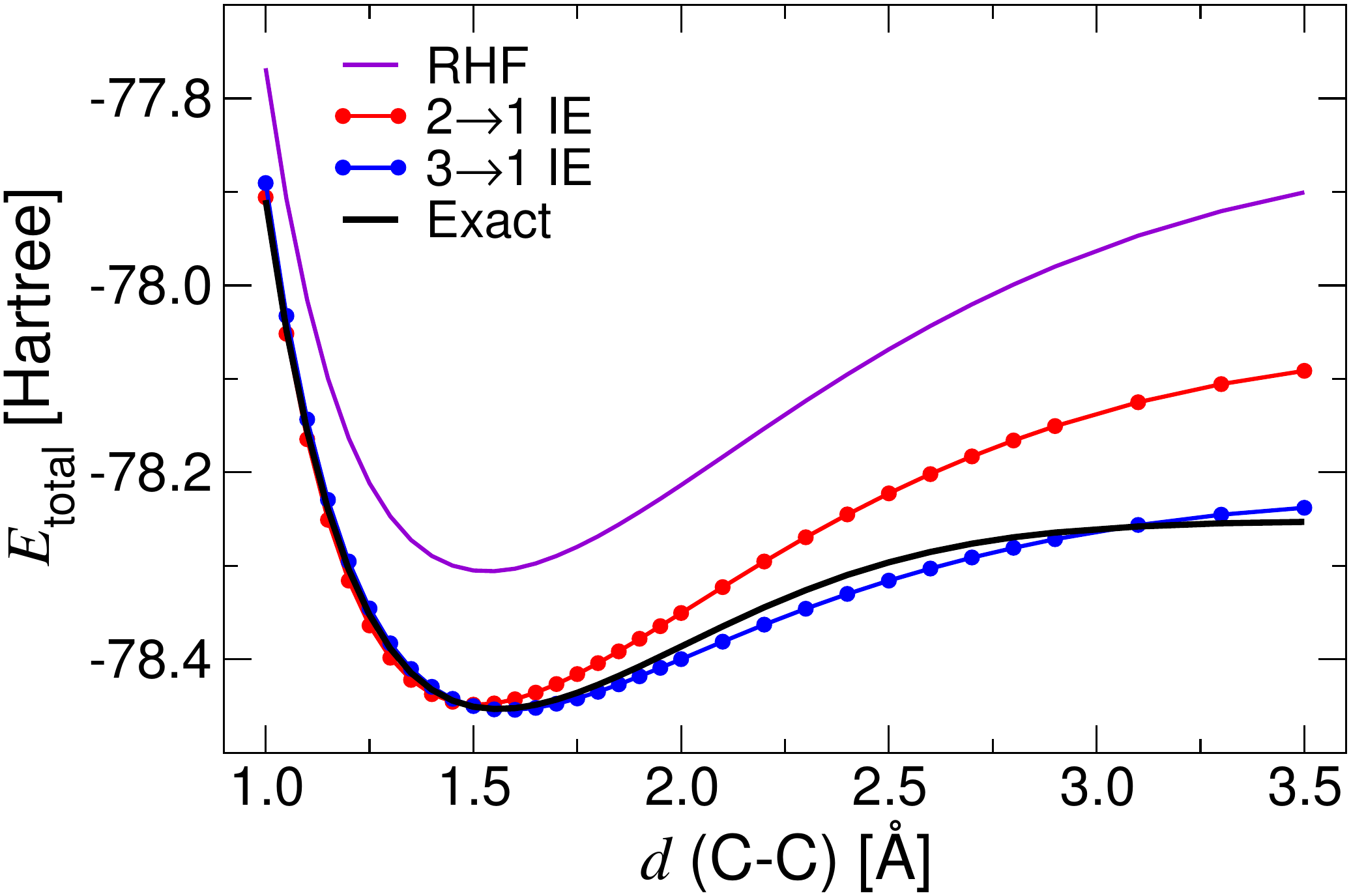}
        \caption{The same calculation as shown in FIG.\ \ref{fig:c2h6_nsc}, with site densities and pair-densities self-consistently determined by IE.}
        \label{fig:c2h6_sc}
    \end{figure*}

    The performance of non-self-consistent IE -- especially the under-correlation at equilibrium geometry at the $2 \to 1$ level -- is an indication that the error of this method is not purely density-driven. To confirm this inference, we repeat the IE calculations above but self-consistently determine the site densities and pair-densities. The results are presented in FIG.\ \ref{fig:c2h6_sc}. By comparing it to FIG.\ \ref{fig:c2h6_nsc}, one can clearly see that imposing self-consistency significantly improves the results at equilibrium geometry at the $2 \to 1$ level, and keeps the high accuracy of $3 \to 1$ IE at the same time. In the dissociation limit, however, imposing self-consistency has opposite effects: the PESs are shifted upwards slightly at both levels (though $3 \to 1$ has a much smaller amplitude) compared to the non-self-consistent results, making the under-correlation problem even more severe therein.
    A scrutiny on the comparison of the site densities and pair-densities obtained by all these methods (Figure S1 in Supporting Information) shows that the change from FIG.\ \ref{fig:c2h6_nsc} to FIG.\ \ref{fig:c2h6_sc} is not density-driven, as the self-consistent densities and pair-densities are consistently worse than the non-self-consistent counterparts. Nevertheless, the results of $3 \to 1$ IE seem to be stable, especially near equilibrium geometries. In those cases, one can safely abandon the self-consistency condition without losing much accuracy.

    \subsection{Correlation Energies at Equilibrium Geometry}   \label{subsec:correlation_energies}

    \begin{figure*}[!h]
        \centering
        \includegraphics[width=0.31\linewidth]{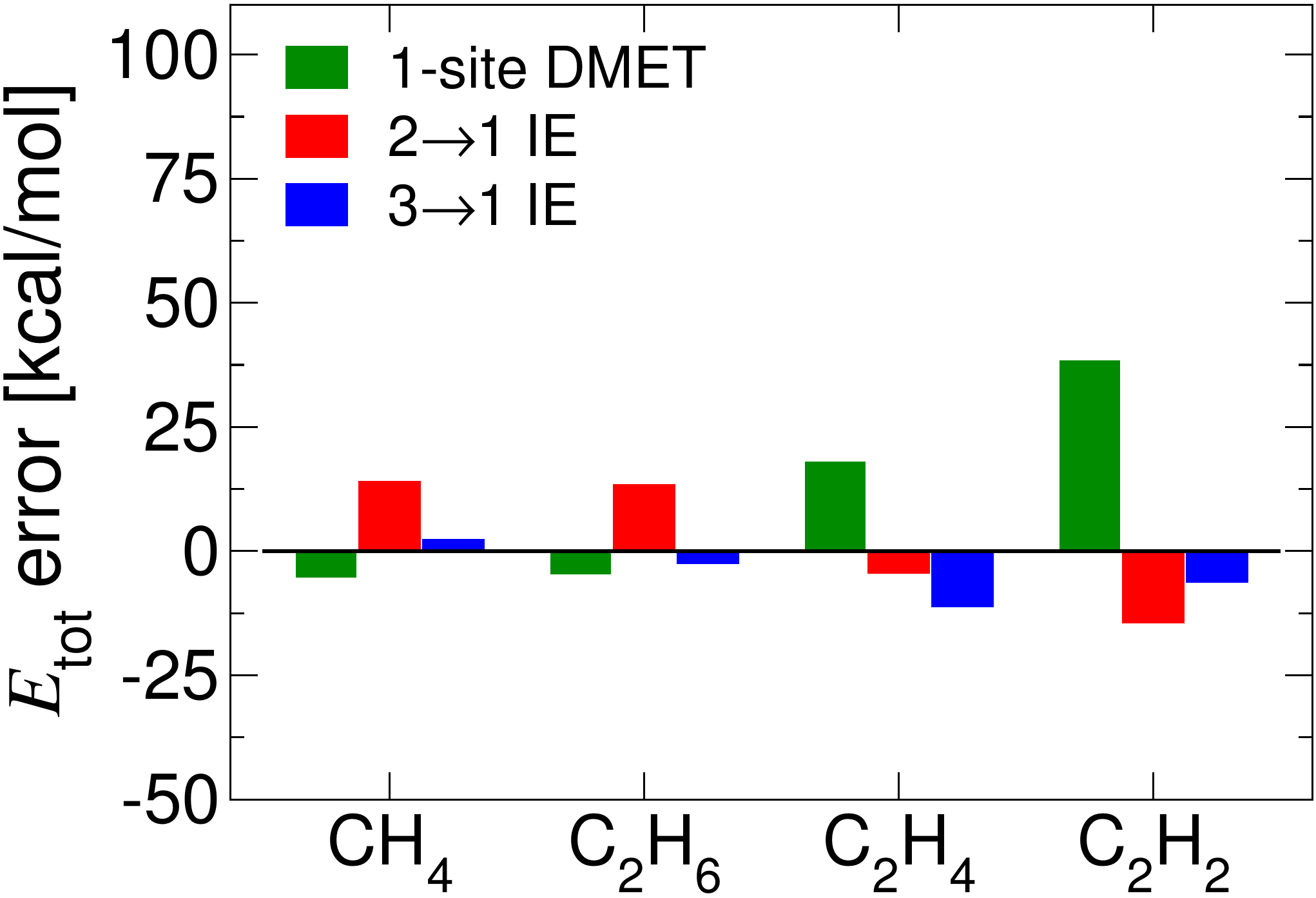}\hfill
        \includegraphics[width=0.31\linewidth]{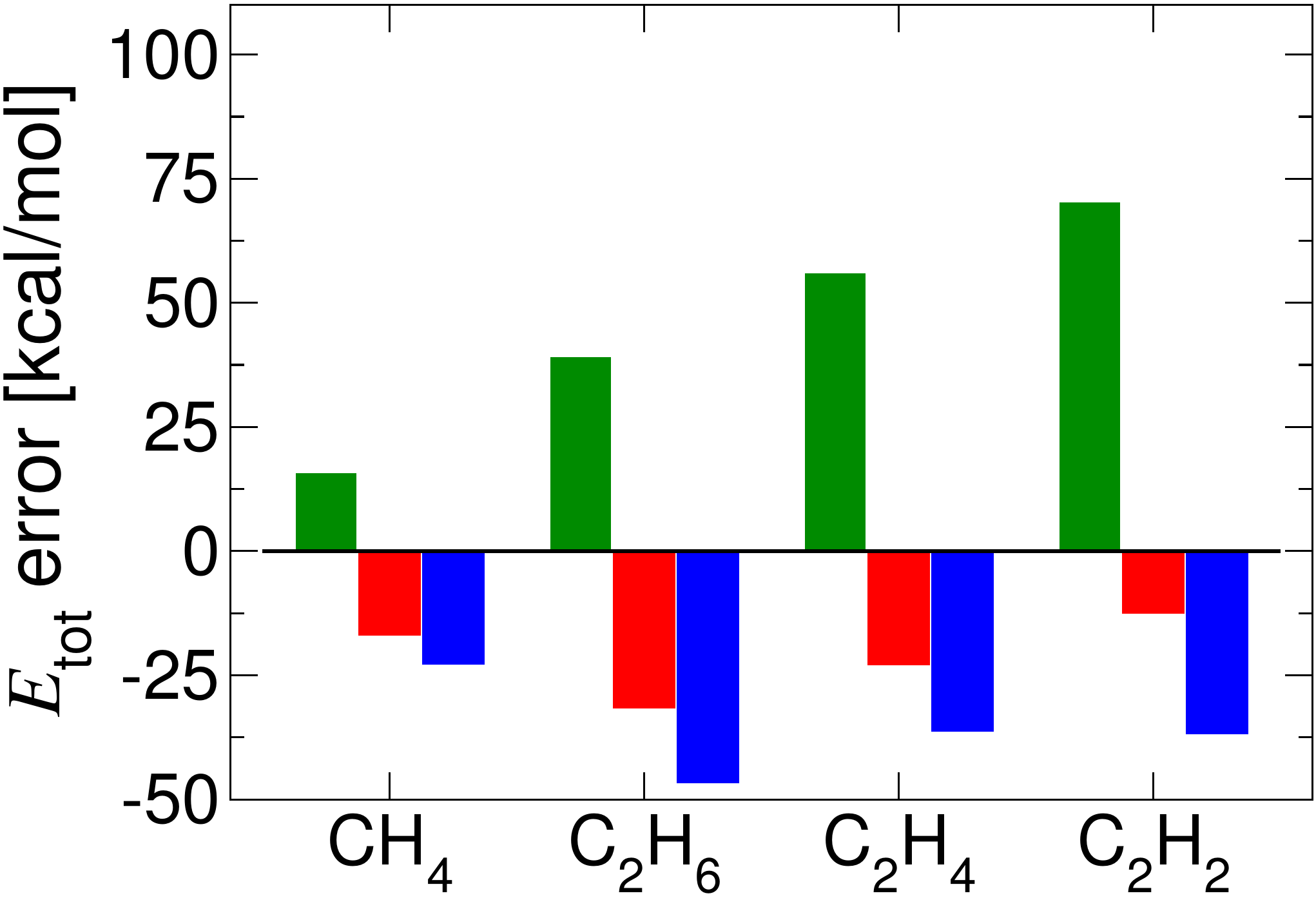}\hfill
        \includegraphics[width=0.31\linewidth]{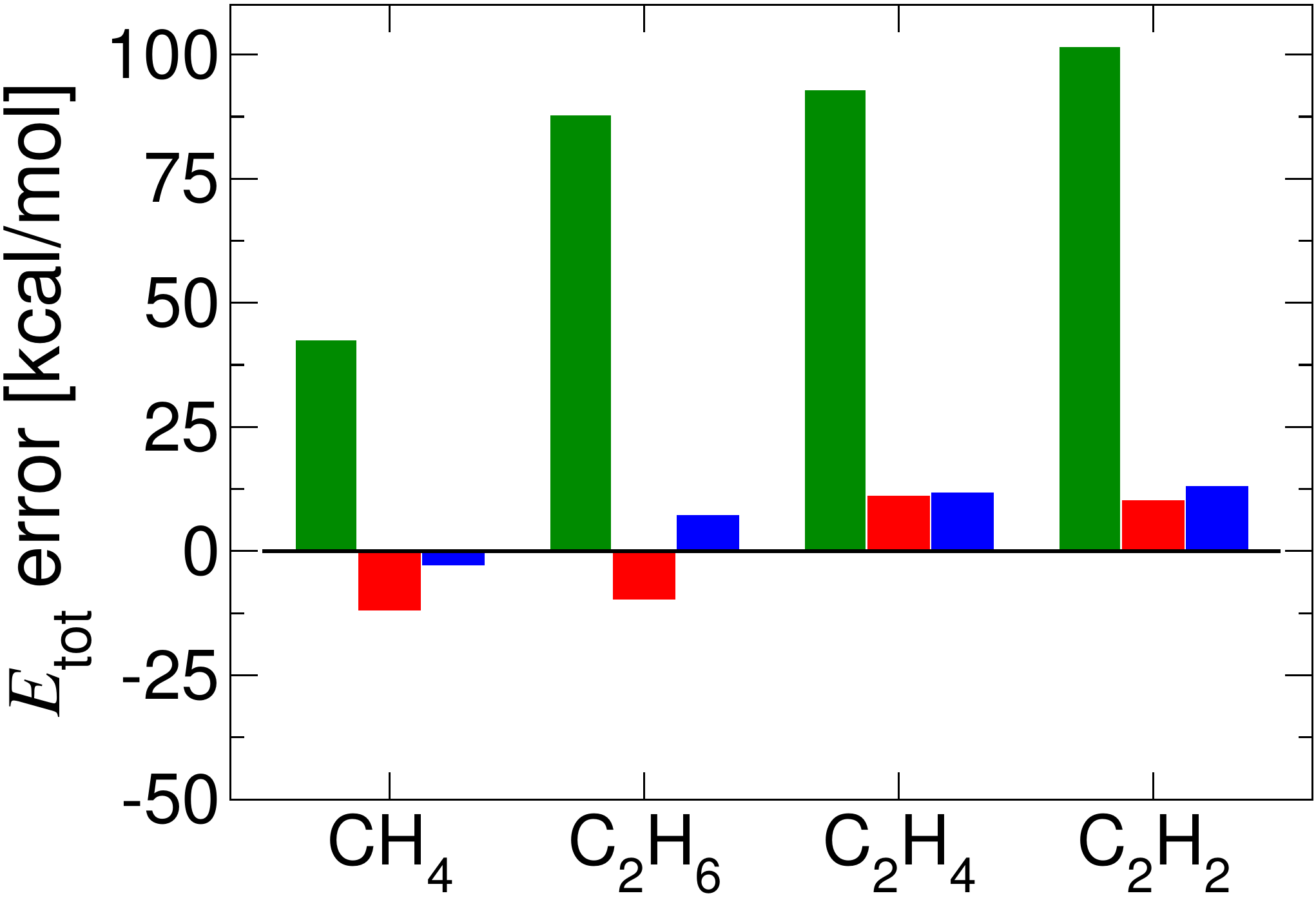}
        \caption{Equilibrium geometry total energy errors (in kcal/mol) obtained by the non-self-consistent IE for several molecules in different basis sets: (a) STO-3G, (b) 3-21G and (c) 6-311G. One-site DMET results are also included for comparison.}
        \label{fig:total_error}
    \end{figure*}

    As a final example, we investigate the effect of larger basis sets. As mentioned in Sec.\ \ref{subsec:Schmidt_decomposition}, any deviation from half-filling deteriorates the performance of Schmidt-space embeddings using HF bath wave functions. In practice, however, large basis sets are often essential to recover the dynamic part of electron correlation. It is thus of significant importance to examine how IE behaves in large basis sets. In FIG.\ \ref{fig:total_error}, we present in terms of bar plot the error of total energies for four molecules at equilibrium geometries predicted by non-self-consistent IE, along with one-site DMET for comparison. Three basis sets of increasing size are used: STO-3G, 3-21G and 6-311G. For the minimal basis, all molecules are close to being half-filled. IE shows a consistent improvement with fragment size in all cases except for \ce{C2H4}, and the errors of the $3 \to 1$ energies are within several kcal/mol's.
    One-site DMET, on the other hand, gives good results for \ce{CH4} and \ce{C2H6} (as already seen in FIG.\ \ref{fig:c2h6_nsc}) but under-correlates badly for the other two unsaturated molecules. Since the densities are the same for all three cases, these results again confirm our conjecture that the errors of Schmidt-space embedding methods are not density-driven: compared to one-site DMET, IE successfully captures the entanglement among different sites, especially when fragments of larger size are used.

    When the basis size is increased, the deviation from half-filling renders the under-correlation problem of one-site DMET even worse, as can be seen from the increasing heights of the green bars in FIG.\ \ref{fig:total_error}b and c. In terms of absolute values, one can see clearly that this is because the correlation energies recovered by one-site DMET remain unchanged or even drop slightly as the basis size gets larger, while the exact correlation energies always go up (Figure S2 in Supporting Information). The same phenomenon appears in IE, but the trend there is more complicated. At first, switching from minimal to double-zeta basis leads to overestimation of correlation energies for all molecules. Moreover, the convergence with fragment size gets reversed: $3 \to 1$ predicts more negative numbers than $2 \to 1$ does, making it worse by going to a higher level of theory.
    Moving further to the triple-zeta basis, however, sets IE back on track: both $2 \to 1$ and $3 \to 1$ energies are of high accuracy. This occurs as a consequence of error cancellation: the correlation energies recovered by IE do not increase when going from 3-21G to 6-311G, which happens to cancel the over-correlation errors in the double-zeta basis accidentally (Figure S2 in Supporting Information). This phenomenon, observed in both one-site DMET and IE, is in fact related to the basis unentanglement problem of the RHF bath. In next section, we will discuss this problem more thoroughly using a specific example.

    \section{Discussion}    \label{sec:discussion}

    \begin{figure}[!h]
        \centering
        \includegraphics[width=1.00\linewidth]{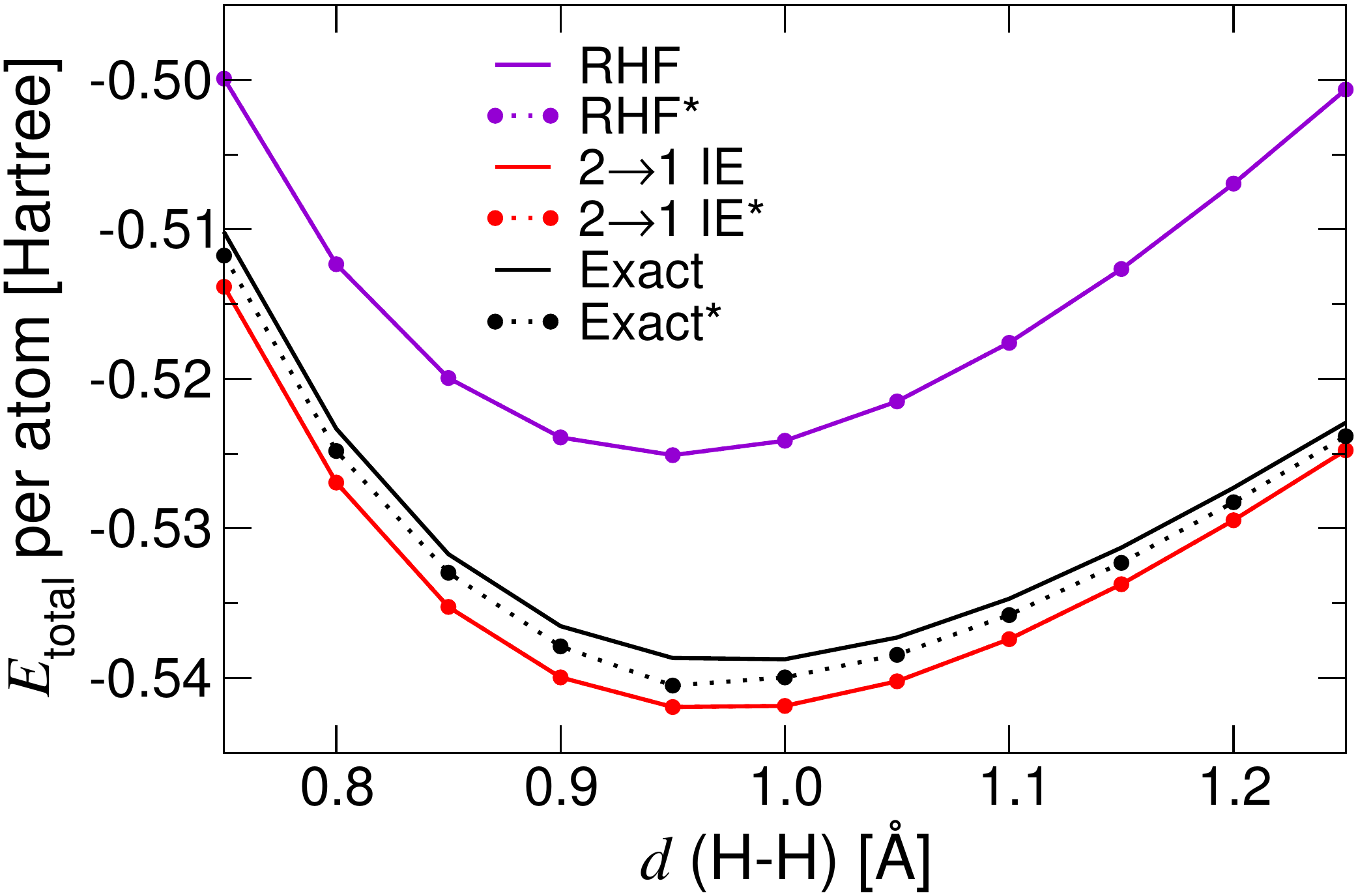}
        \caption{A re-plot of FIG.\ \ref{fig:h10} for the total energy of \ce{H_{10}} near equilibrium geometry. Data labelled with an asterisk are computed in the new basis STO-3G$^*$.}
        \label{fig:h10_sto-3gstar}
    \end{figure}

    In Sec.\ \ref{subsec:Schmidt_decomposition} we briefly mentioned that the unentanglement in bath wave function (in this work, RHF) leads to linear dependency in the Schmidt decomposition [eqn (\ref{eq:Schmidt_decomposition_general})] and effectively degrades the basis. Here we present a specific example that embodies this problem.
    We repeat our calculations in Sec.\ \ref{subsec:hydrogen_ring} using a homemade basis set STO-3G$^*$, obtained by adding one $p_z$ orbital to each STO-3G hydrogen (assuming the atomic ring lies in the $xy$ plane). The PESs obtained in the new basis are presented in FIG.\ \ref{fig:h10_sto-3gstar} along with the original results for comparison. The first thing to notice is that the RHF solution remains unchanged. Population analysis suggests that the set of $p$ orbitals are not occupied at all. This is expected by our construction of the new basis since all of the one-electron atomic integrals between the new orbitals and the original $1s$ orbitals vanish by symmetry. The only non-vanishing parts are the two-electron integrals involving an even number of $p$ orbitals such as $(p_zp_z|1s 1s)$. Unfortunately, these non-trivial interactions are not captured by the mean-field wave function due to it being a one-electron theory.
    As a result, the Schmidt space in the new basis is exactly the same as in the original one, which explains the concurrence of two IE curves in FIG.\ \ref{fig:h10_sto-3gstar}. For the exact solution, however, those interactions do contribute to the total energy, and one can see a lower energy in the new basis.

    This specifically designed example helps us to understand the trend shown in FIG.\ \ref{fig:total_error}. As mentioned in Sec.\ \ref{subsec:correlation_energies}, in terms of absolute values, the correlation energies recovered by both one-site DMET and IE cease to increase once the basis set reaches a certain size (see also Figure S2 in Supporting Information). This observation indicates that the basis unentanglement problem of the mean-field bath could be common in realistic systems, especially when large-sized bases are used.

    \section{Conclusion}    \label{sec:conclusion}

    In conclusion, we have introduced Incremental Embedding, a new Schmidt-space fragment embedding scheme that allows the combination of arbitrary overlapping fragments without the knowledge of edge sites. Underlying this new method is one of the key concept introduced in this work, Schmidt reduction, which allows information to be encoded from a large embedding space to a smaller one. Based on this technique, IE constructs one-site effective Hamiltonians for all sites by hierarchically incorporating corrections from embeddings involving two-site fragments, three-site fragments, and \latin{etc} to the mean-field approximation.
    The potential double counting problem is avoided by an elaborate application of the method of increments. This method can be viewed as an unambiguous many-site generalization of one-site DMET. It can be made either self-consistent or non-self-consistent. The computational scaling is $O(K^m)$ for the lowest few levels in the hierarchy, which are much lower than most correlated wave function theories.

    Numerical simulations on small molecules in atom-centered Gaussian bases suggest that the convergence with fragment size is quick in small bases; most of the electron correlation is recovered for all molecules tested, even when truncated at the $3 \to 1$ level. Imposing self-consistency in site densities and pair-densities improves the performance of IE considerably near equilibrium geometry, through an approach that is not density-driven. For larger bases, both IE and one-site DMET recover only a fraction of the correlation energy, which can be attributed to the more general basis unentanglement problem of the RHF bath. In summary, this work marks the first attempt of applying Schmidt-space embedding methods to realistic molecular systems using overlapping fragments.

    In the future, IE can be extended in a number of directions. First of all, up to this point we restrict ourselves to FCI for the embedding Hamiltonian, which is computationally expensive and can only be applied to fragments of limited size. One can, of course, pursue other high-level solvers such as DMRG and CCSD, which have better computational scaling and can therefore be extended to fragments of larger size. Second, as for the site basis, we restrict ourselves in this work to SOAOs or LMOs by the Foster-Boys scheme for the sake of simplicity, but other choices do exist. In analogy to local correlation methods such as local MP2\cite{Pulay83CPL,Pulay86TCA,Saebo87JCP} and local CCSD\cite{Hampel96JCP}, perhaps the most straightforward way is to explore the possibilities of using other LMOs such as those given by the Pipek-Mezey scheme\cite{Pipek89JCP,Boughton93JCC} and the Edmiston-Ruedenberg scheme\cite{Edmiston63RMP}.
    Last but not least, the conflict between the half-filled embedding space and the maximum number of entangled sites in a HF bath calls for a better bath wave function. In this regard, the Hartree-Fock-Bogoliubov\cite{Bogoliubov58FP,Bogoliubov59SPU} (HFB) wave function might be a good candidate because it is (i) always half-filled in the quasi-particle space, and (ii) still a mean-field theory and therefore retains the simplicity of embedding Hamiltonians in a mean-field bath.

    \begin{acknowledgements}
        HY thanks Dr.\ Tianyu Zhu for the discussion on the method of increments. This work was funded by a grant from the NSF (Grant No.\ CHE-1464804). TV is a David and Lucille Packard Foundation Fellow.
    \end{acknowledgements}

\bibliography{refs}

\end{document}